\documentclass[12pt]{article}
\usepackage{graphicx}
\usepackage{lineno}
\usepackage{xspace}
\usepackage{placeins}
\usepackage[colorlinks = true,
            linkcolor = blue,
            urlcolor  = blue,
            citecolor = blue,
            anchorcolor = blue]{hyperref}

\newcommand\pubdate{\today}

\def\Title#1{\begin{center} {\Large #1 } \end{center}}
\def\Author#1{\begin{center}{ \sc #1} \end{center}}
\def\Address#1{\begin{center}{ \it #1} \end{center}}

\newcommand\pubblock{\rightline{\begin{tabular}{l}  \\ 
         \pubdate  \end{tabular}}}
\newenvironment{Abstract}{\begin{quotation}  }{\end{quotation}}
\newenvironment{Presented}{\begin{quotation} \begin{center} 
             PRESENTED AT\end{center}\bigskip 
      \begin{center}\begin{large}}{\end{large}\end{center} \end{quotation}}

\newcommand*{\ttbar}{\ensuremath{t\bar{t}}\xspace}
\newcommand*{\ttj}{\ensuremath{t\bar{t}j}\xspace}

\def\beq{\begin{equation}}
\def\eeq#1{\label{#1}\end{equation}}
\def\eeqn{\end{equation}}

\newenvironment{Eqnarray}%
   {\arraycolsep 0.14em\begin{eqnarray}}{\end{eqnarray}}
\def\beqa{\begin{Eqnarray}}
\def\eeqa#1{\label{#1}\end{Eqnarray}}
\def\eeqan{\end{Eqnarray}}

\let\bar=\overbar

\def\lsim{\mathrel{\raise.3ex\hbox{$<$\kern-.75em\lower1ex\hbox{$\sim$}}}}
\def\gsim{\mathrel{\raise.3ex\hbox{$>$\kern-.75em\lower1ex\hbox{$\sim$}}}}

\def\del{\partial}
\def\Dslash{\not{\hbox{\kern-4pt $D$}}}
\def\dslash{\not{\hbox{\kern-2pt $\del$}}}
\def\pslash{\not{\hbox{\kern-2pt $p$}}}
\def\ETmiss{\not{\hbox{\kern-4pt $E$}}_T}

\def\Dlr{\mathrel{\raise1.5ex\hbox{$\leftrightarrow$\kern-1em\lower1.5ex\hbox{$D$}}}}

\def\MSB{{\bar{M \kern -2pt S}}}
\def\msb{{\bar{\scriptsize M \kern -1pt S}}}

\def\drb{{\bar{\scriptsize D \kern -1pt R}}}

\def\TeV{{\rm TeV}\xspace}

\begin{document}


\begin{titlepage}
 \pubblock
\vfill
\Title{Future Top Quark Pole Mass Improvements from PDF Updates}
\vfill
\Author{Jason Gombas-Salazar, Reinhard Schwienhorst, Jarrett Fein,  Sara Sawford}
\Address{Michigan State University, East Lansing, MI, United States of America}
\vfill
\begin{Abstract}
The dependence of the top-quark mass measurement in top-quark pair production on the parton distribution functions (PDF) is explored through differential mass distributions in $t\bar{t}$
 and $t\bar{t}j$ production at the LHC and a future 100 TeV proton-proton collider. The top-quark mass uncertainty is obtained from chi-squared fits to invariant mass distributions from simulations assuming different top pole masses around the nominal value of 172.5 GeV. The PDF uncertainties of the differential distributions are used in the chi-square evaluation and reduced through a fit to differential distributions in $t\bar{t}$ and $t\bar{t}j$ production.
\end{Abstract}
\vfill
\begin{Presented}
DIS2023: XXX International Workshop on Deep-Inelastic Scattering and
Related Subjects, \\
Michigan State University, USA, 27-31 March 2023 \\
     \includegraphics[width=9cm]{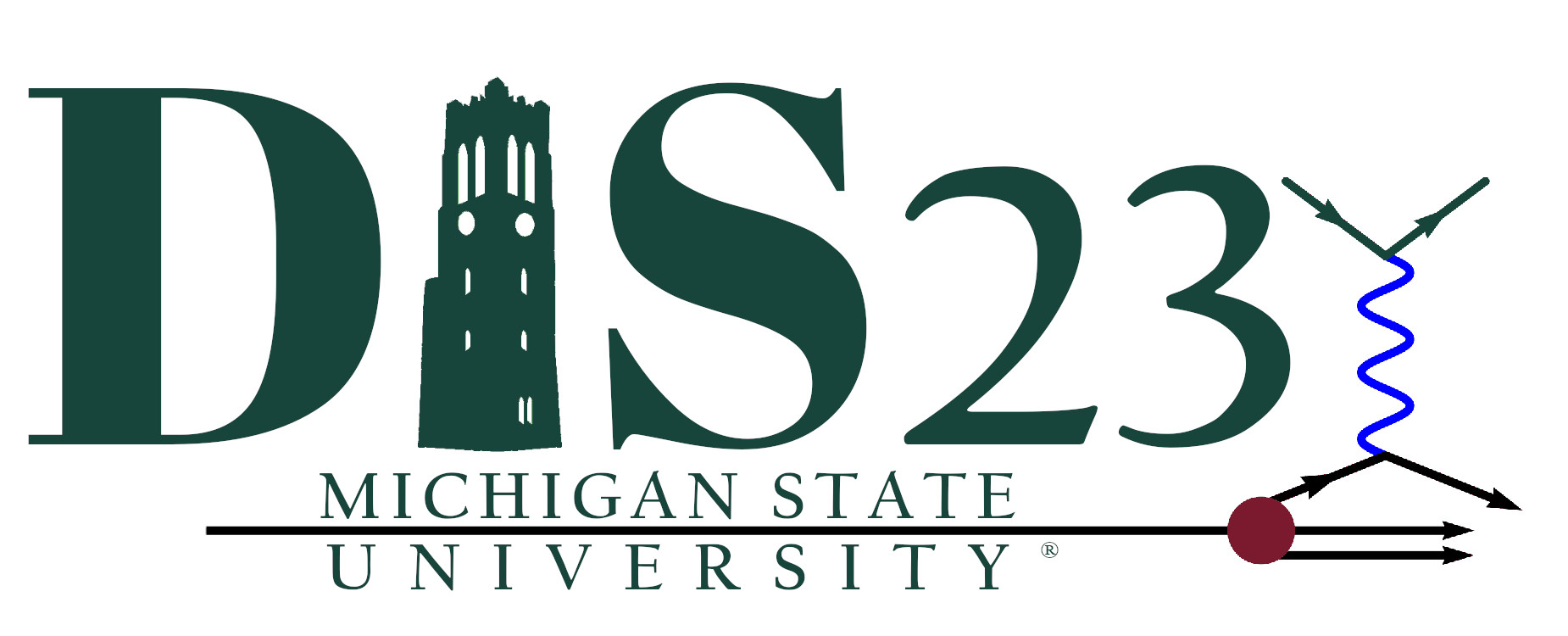}
\end{Presented}
\vfill
\small{}
\end{titlepage}

\section{Introduction}
The top quark mass is an important measurement because it determines if we live in a stable, unstable, or meta-stable universe~\cite{Degrassi:2012ry}. It is also a key ingredient to global electroweak fits~\cite{Haller:2018nnx}. Future measurements of the top quark mass are limited by theoretical uncertainties more than experimental ones~\cite{Schwienhorst:2022yqu,Hoang:2020iah}. 

The top quark pole mass is measured by unfolding differential top-quark pair (\ttbar) event distributions to the parton level~\cite{ATLAS:2017dhr,CMS:2019esx} and then comparing those distributions directly to precise QCD calculations~\cite{Guzzi:2014wia,Czakon:2011xx}. While this is theoretically well-defined, it has large theoretical uncertainties. In particular the uncertainty due to the parton-distribution functions (PDFs) contributes significantly to the top quark pole mass. For the inclusive cross-section calculation, the PDFs are the dominant uncertainty~\cite{Czakon:2013goa}. An alternative method in measuring the top quark pole mass relies on top-quark pair events that contain an additional jet (\ttj), which is free from some theoretical issues in describing the threshold region of \ttbar production~\cite{Alioli:2013mxa,Alioli:2022ttk}.

In the presented study, we explore measurements of the top-quark pole mass in \ttbar and in \ttj production at the parton level, focusing on the contribution from parton distribution functions. We consider two different processes separately. 
\begin{itemize}
    \item For \ttbar production, we fit the top-quark mass to the differential distribution of the invariant mass of the \ttbar system. 
    \item For \ttj production, we fit the top-quark mass to the differential distribution of the $\rho$ variable, which is proportional to the inverse of the invariant mass of the \ttj system.
\end{itemize}
In both mass extractions, we evaluate chi-square based on the PDF uncertainty, which can be reduced through a fit to kinematic variables using ePump~\cite{Kadir:2020yml,Schmidt:2018hvu}. We study proton-proton colliders in various configurations, including the LHC at 8, 13, 13.6, 14~\TeV~\cite{Evans:2008zzb} and the 100~\TeV proton-proton collider~\cite{FCC:2018byv,FCC:2018vvp,CEPCStudyGroup:2018rmc}.

\section{Simulation Setup}
\begin{figure}[!h]
  \center {\includegraphics[width=0.45\textwidth]{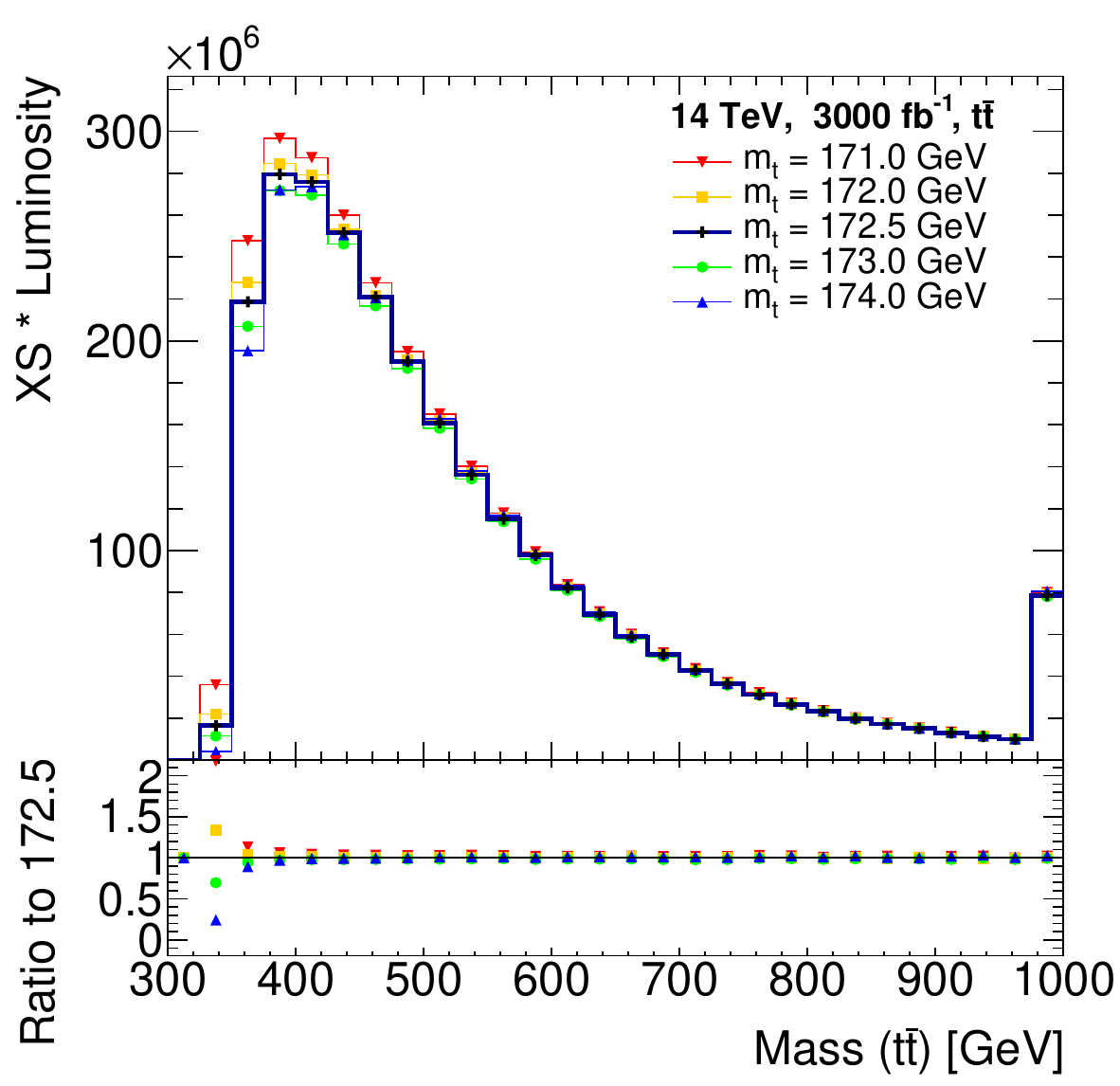}
           \includegraphics[width=0.45\textwidth]{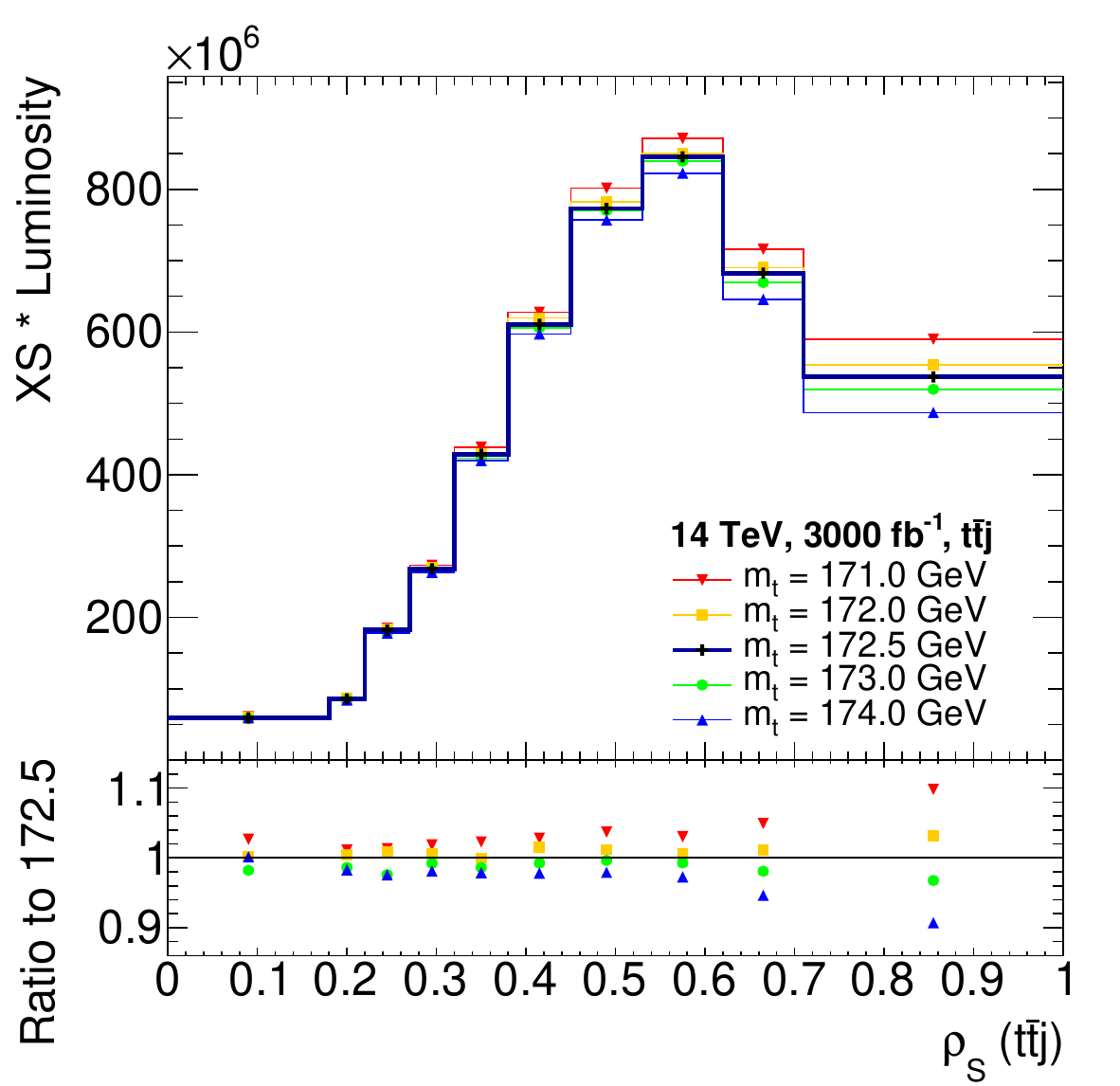}}
  \caption{\\
    Differential distributions are calculated using CT18NLO, a center-of-mass energy of $14~\textrm{TeV}$, and normalized to $3000~\textrm{fb}^{-1}$. The last bin in the distribution is the overflow bin. Each plot shows the calculations assuming different top-quark pole mass hypothesis in the region of the nominal value of $172.5~\textrm{GeV}$. The ratio of each mass hypothesis to the nominal value is also shown here. {[Left]} The differential invariant \ttbar mass distribution. {[Right]} The differential $\rho=340~\textrm{GeV}/m_{\ttj}$ distribution.
    }
  \label{fig:differential_distributions}
\end{figure}

Madgraph~\cite{Alwall:2011uj,Alwall:2014hca} is used to generate top-quark pair production events and top-quark pair production events with an additional jet at next-to-leading order in QCD (NLO). Events don't include next-to-next-to-leading order corrections (NNLO) or resummation effects. CT18NLO was the PDF selected for the Madgraph event generation. Differential distributions of the \ttbar mass distribution and the differential $\rho$ distribution are created from these generated events directly and can be seen in Fig.~\ref{fig:differential_distributions}. Detector simulation, efficiencies and acceptances are not considered in this study. The tops generated directly from Madgraph are not decayed. $\rho$ is chosen as the variable for \ttj because it is the most sensitive variable to the top quark mass for the \ttj process. 

Most of the significant deviations from the nominal hypothesis occur at the \ttbar mass threshold which is twice the mass of the top quark. The most significant deviations in the $\rho$ distribution are in the last bin. Constraining the PDF uncertainty in these regions will most significantly reduce the top quark mass uncertainty. The rapidity of the top quark in \ttbar events and $p_Z$ of the \ttj system for \ttj are the variables we chose to reduce the uncertainty in these regions of phase space.

\section{Pseudo-data and PDF Updating}
The same generated events that are used to create the differential mass distributions are used to calculate the \textit{pseudo-data} distributions. Although this calculation uses the same PDF sets which we intend to constrain, our goal is not to predict how the data will shift the PDF set, but rather study how the new data will impact and constrain the uncertainties. Therefore, the PDF shape will remain the same, as it must, but the uncertainty bands on the PDF will shrink due to the introduced pseudo-data. The pseudo data is normalized to the predicted number of events and the systematic uncertainty is set to 1\%. The 1\% uncertainty is ambitious but should be within reach of the HL-LHC~\cite{ATLAS:2019hxz,CMS:2019esx} though more precise theory predictions are required in addition to precise measurements. 

\begin{figure}[!h]
  \center {\includegraphics[width=0.45\textwidth]{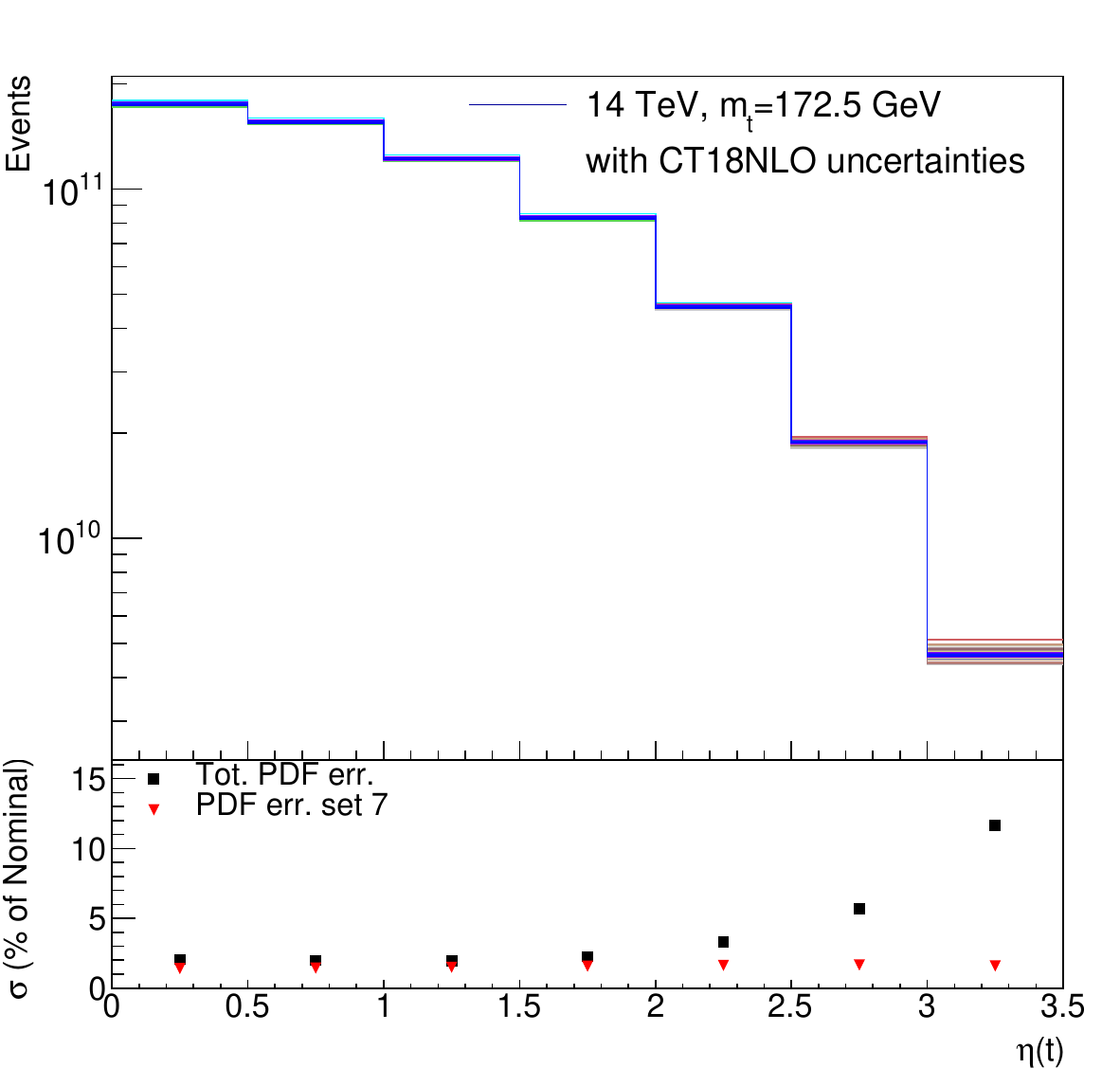}
           \includegraphics[width=0.45\textwidth]{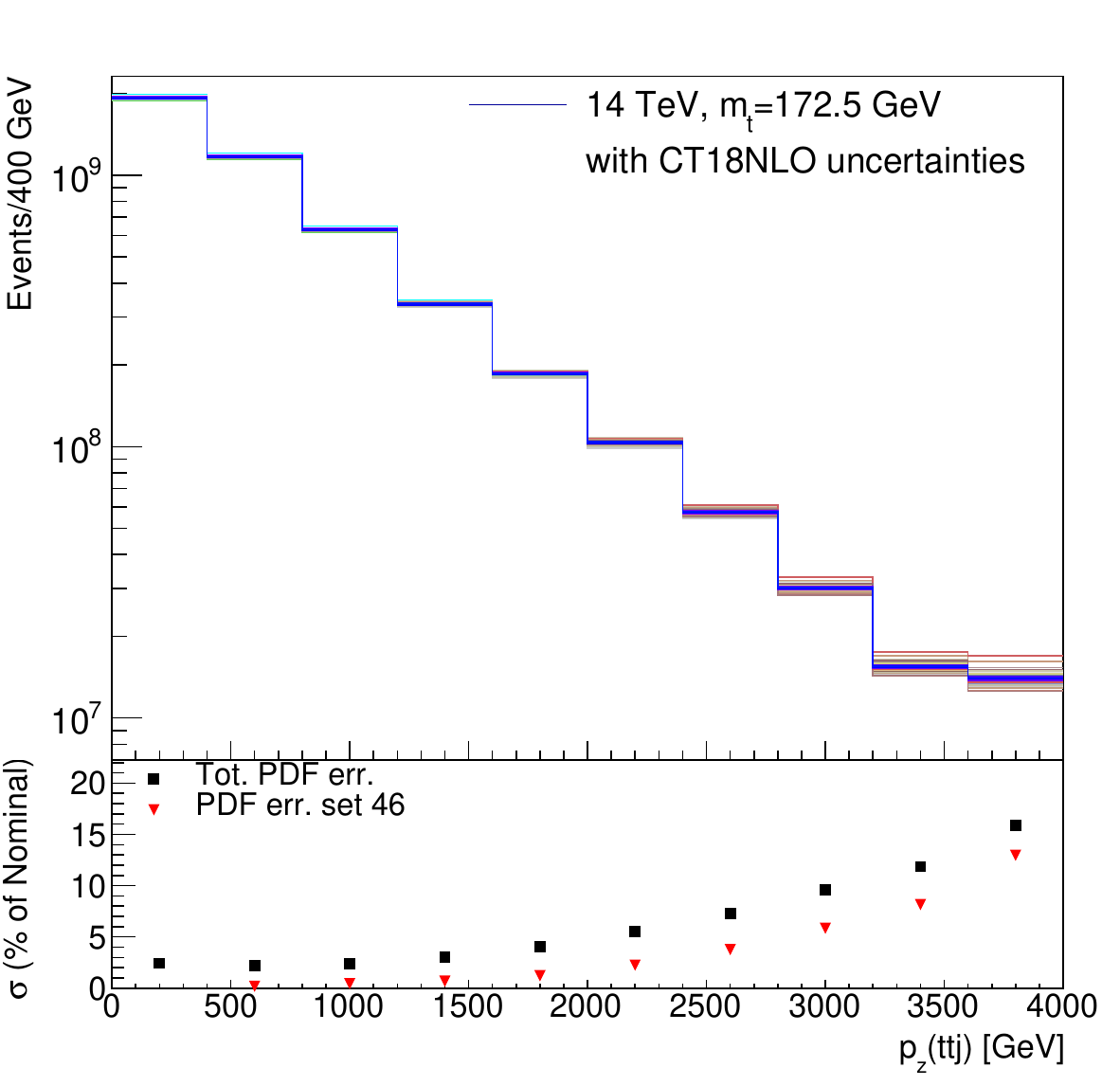}}
  \caption{\\
    The pseudo data that is used in the ePump PDF update. [Left] The rapidity of the top quark distribution in the \ttbar events. [Right] The $p_Z$ of the \ttj system in the \ttj events. 
    }
  \label{fig:measurements}
\end{figure}

The pseudo-data that is input into ePump is shown in Fig.~\ref{fig:measurements}. The rapidity of the top quark is chosen to constrain the PDF uncertainty for the \ttbar process. and the $p_Z$ of the \ttj system is chosen to reduce the PDF uncertainty for the \ttj process.

The uncertainty increases with bigger top quark rapidity and with bigger $p_Z$. This can mostly be attributed to large uncertainties in the gluon PDF at high momentum fraction and high energy scale.  

\begin{figure}[!h]
  \center {\includegraphics[width=0.45\textwidth]{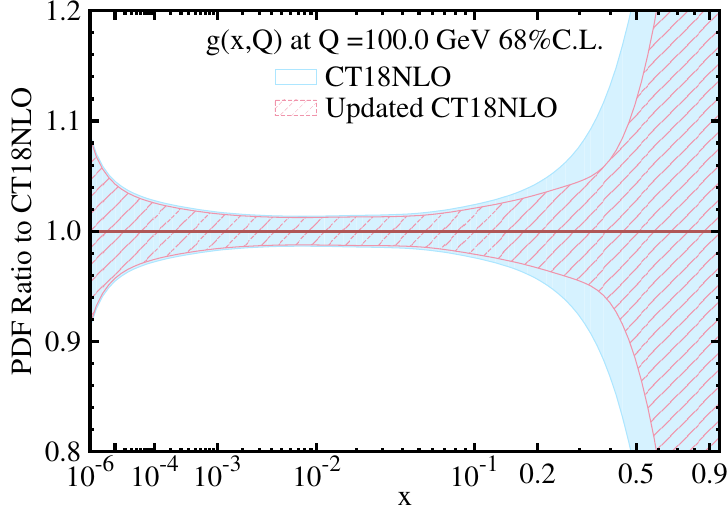}
           \includegraphics[width=0.45\textwidth]{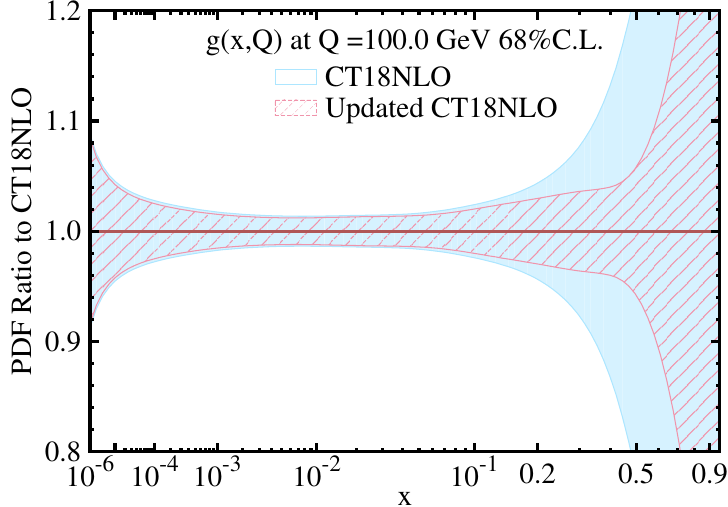}}
  \caption{\\
    Gluon PDF that shows the difference between the original CT18NLO PDF set and the updated PDF set using the pseudo data described above. [Left] The updated PDF set in the \ttbar production set using the rapidity of the top quark. [Right] The updated PDF set in the \ttj production set using the $p_Z$ of the \ttj system. 
    }
  \label{fig:pdf_plots}
\end{figure}

After the ePump update, which uses the Hessian approach,~\cite{Schmidt:2018hvu} the resulting gluon PDF can be seen in Fig.~\ref{fig:pdf_plots}. Other PDFs were looked at and their changes were minimal. Fig.~\ref{fig:pdf_plots} shows that the region that most significantly changes in the gluon PDF is the high momentum fraction region. The uncertainty is given by the eigenvectors of the PDF, following the CTEQ prescription for 68\% CL intervals~\cite{Hou:2019efy}.

This updated PDF set, with its reduced uncertainties, is then used to calculate chi-square again to compare with the original PDF set. 

\section{Results}
\begin{figure}[!h]
  \center {\includegraphics[width=0.45\textwidth]{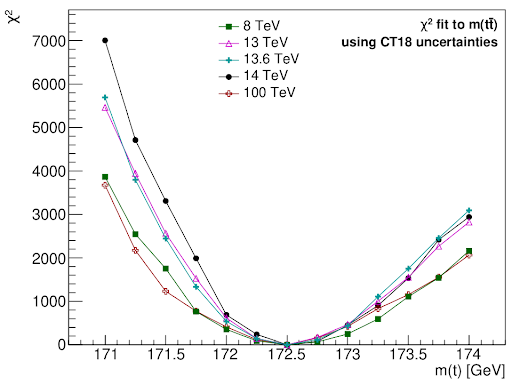}
           \includegraphics[width=0.45\textwidth]{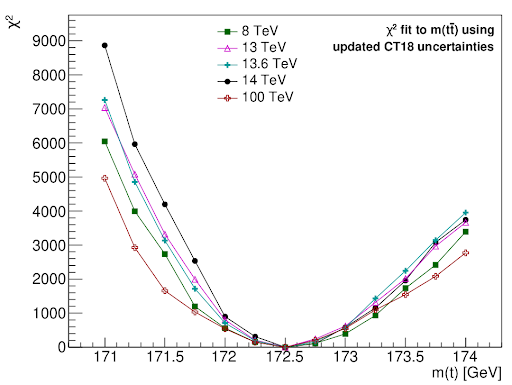}}
  \caption{\\
    Chi-square calculations for \ttbar for different proton-proton collider energies.
    }
  \label{fig:chi2ttbarall}
\end{figure}

\begin{figure}[!h]
  \center {\includegraphics[width=0.45\textwidth]{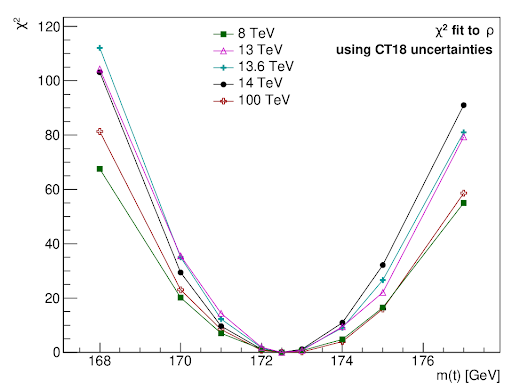}
           \includegraphics[width=0.45\textwidth]{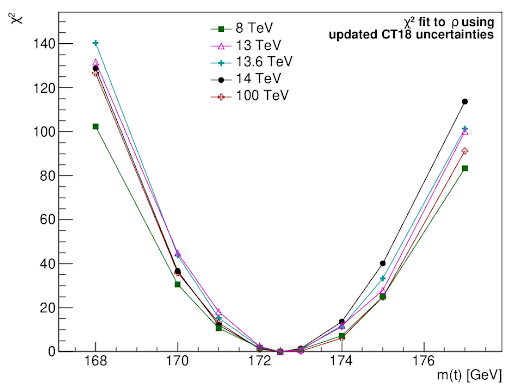}}
  \caption{\\
    Chi-square calculations for \ttj for different proton-proton collider energies.
    }
  \label{fig:chi2ttjall}
\end{figure}

The predicted impact the introduced pseudo data on the top quark pole mass from the reduced PDF uncertainties is shown in Fig.~\ref{fig:chi2ttbarall} for \ttbar and in Fig.~\ref{fig:chi2ttjall} for \ttj. We use the equation $\chi^2=\Sigma(O_i-E_i)^2/\sigma_i$ for our calculations, where $i$ indicates the bin in the histogram, $O_i$ is the pseudo data, $E_i$ is the nominal expected value, and $\sigma_i$ is the PDF uncertainty. For any given energy, the chi-square increases with each update which is expected. The skew of the \ttbar chi-square plots can be explained by large asymmetry of the invariant \ttbar mass distribution. 

\section{Conclusions}
We have presented a study of the top-quark mass in top-quark pair production and top-quark pair plus an additional jet production at hadron colliders using the invariant mass of the \ttbar system in \ttbar production and $\rho$ in \ttj production as the sensitive variable. We study the LHC at 8, 13, 13.6, and 14~\TeV, and the future 100~TeV proton-proton collider. We explore the sensitivity of the top-quark mass fit to parton distribution functions and evaluate the uncertainty on the top-quark mass due to the PDF uncertainty with the CT18NLO PDF set. This is done separately for the \ttbar and \ttj production case. The PDF uncertainty can be improved by fitting the distribution of the rapidity of the top quark, which improves the top-quark mass uncertainty by about 12\%-15\%. The PDF uncertainty can also be improved by fitting the distribution of the longitudinal momentum $p_Z$, which improves the top-quark mass uncertainty by about 13\%-20\%. Further improvement of the PDF impact on the top-quark mass measurements in production require additional input from outside top-quark pair production.

\bigskip
\noindent
\textbf{Funding information} This project was supported in part by the National Science Foundation under grant no. PHY-2012165

\bibliographystyle{JHEP}
\bibliography{refs}  

\providecommand{\href}[2]{#2}\begingroup\raggedright\begin{thebibliography}{10}

\bibitem{Degrassi:2012ry}
G.~Degrassi, S.~Di~Vita, J.~Elias-Miro, J.R.~Espinosa, G.F.~Giudice, G.~Isidori
  et~al., \emph{{Higgs mass and vacuum stability in the Standard Model at
  NNLO}}, \href{https://doi.org/10.1007/JHEP08(2012)098}{\emph{JHEP} {\bfseries
  08} (2012) 098} [\href{https://arxiv.org/abs/1205.6497}{{\ttfamily
  1205.6497}}].

\bibitem{Haller:2018nnx}
J.~Haller, A.~Hoecker, R.~Kogler, K.~M\"onig, T.~Peiffer and J.~Stelzer,
  \emph{{Update of the global electroweak fit and constraints on
  two-Higgs-doublet models}},
  \href{https://doi.org/10.1140/epjc/s10052-018-6131-3}{\emph{Eur. Phys. J. C}
  {\bfseries 78} (2018) 675}
  [\href{https://arxiv.org/abs/1803.01853}{{\ttfamily 1803.01853}}].

\bibitem{Schwienhorst:2022yqu}
{R.~Schwienhorst and D.~Wackeroth (editors)}, \emph{{Report of the Topical
  Group on Top quark physics and heavy flavor production for Snowmass 2021}},
  \href{https://arxiv.org/abs/2209.11267}{{\ttfamily 2209.11267}}.

\bibitem{Hoang:2020iah}
A.H.~Hoang, \emph{{What is the Top Quark Mass?}},
  \href{https://doi.org/10.1146/annurev-nucl-101918-023530}{\emph{Ann. Rev.
  Nucl. Part. Sci.} {\bfseries 70} (2020) 225}
  [\href{https://arxiv.org/abs/2004.12915}{{\ttfamily 2004.12915}}].

\bibitem{ATLAS:2017dhr}
{ATLAS Collaboration}, \emph{{Measurement of lepton differential distributions
  and the top quark mass in $t\bar{t}$ production in $pp$ collisions at
  $\sqrt{s}=8$ TeV with the ATLAS detector}},
  \href{https://doi.org/10.1140/epjc/s10052-017-5349-9}{\emph{Eur. Phys. J. C}
  {\bfseries 77} (2017) 804}
  [\href{https://arxiv.org/abs/1709.09407}{{\ttfamily 1709.09407}}].

\bibitem{CMS:2019esx}
{CMS Collaboration}, \emph{{Measurement of $\mathrm{t\bar t}$ normalised
  multi-differential cross sections in pp collisions at $\sqrt s=13$ TeV, and
  simultaneous determination of the strong coupling strength, top quark pole
  mass, and parton distribution functions}},
  \href{https://doi.org/10.1140/epjc/s10052-020-7917-7}{\emph{Eur. Phys. J. C}
  {\bfseries 80} (2020) 658}
  [\href{https://arxiv.org/abs/1904.05237}{{\ttfamily 1904.05237}}].

\bibitem{Guzzi:2014wia}
M.~Guzzi, K.~Lipka and S.-O.~Moch, \emph{{Top-quark pair production at hadron
  colliders: differential cross section and phenomenological applications with
  DiffTop}}, \href{https://doi.org/10.1007/JHEP01(2015)082}{\emph{JHEP}
  {\bfseries 01} (2015) 082} [\href{https://arxiv.org/abs/1406.0386}{{\ttfamily
  1406.0386}}].

\bibitem{Czakon:2011xx}
M.~Czakon and A.~Mitov, \emph{{Top++: A Program for the Calculation of the
  Top-Pair Cross-Section at Hadron Colliders}},
  \href{https://doi.org/10.1016/j.cpc.2014.06.021}{\emph{Comput. Phys. Commun.}
  {\bfseries 185} (2014) 2930}
  [\href{https://arxiv.org/abs/1112.5675}{{\ttfamily 1112.5675}}].

\bibitem{Czakon:2013goa}
M.~Czakon, P.~Fiedler and A.~Mitov, \emph{{Total Top-Quark Pair-Production
  Cross Section at Hadron Colliders Through $O(\alpha^4_S)$}},
  \href{https://doi.org/10.1103/PhysRevLett.110.252004}{\emph{Phys. Rev. Lett.}
  {\bfseries 110} (2013) 252004}
  [\href{https://arxiv.org/abs/1303.6254}{{\ttfamily 1303.6254}}].

\bibitem{Alioli:2013mxa}
S.~Alioli, P.~Fernandez, J.~Fuster, A.~Irles, S.-O.~Moch, P.~Uwer et~al.,
  \emph{{A new observable to measure the top-quark mass at hadron colliders}},
  \href{https://doi.org/10.1140/epjc/s10052-013-2438-2}{\emph{Eur. Phys. J. C}
  {\bfseries 73} (2013) 2438}
  [\href{https://arxiv.org/abs/1303.6415}{{\ttfamily 1303.6415}}].

\bibitem{Alioli:2022ttk}
S.~Alioli, J.~Fuster, M.V.~Garzelli, A.~Gavardi, A.~Irles, D.~Melini et~al.,
  \emph{{Top-quark mass extraction from $t\bar{t}j +X$ events at the LHC:
  theory predictions}},  in \emph{{2022 Snowmass Summer Study}}, 3, 2022
  [\href{https://arxiv.org/abs/2203.07344}{{\ttfamily 2203.07344}}].

\bibitem{Kadir:2020yml}
M.~Kadir, A.~Ablat, S.~Dulat, T.-J.~Hou and I.~Sitiwaldi, \emph{{The impact of
  ATLAS and CMS single differential top-quark pair measurements at $\sqrt {s}$
  = 8 TeV on CTEQ-TEA PDFs}},
  \href{https://doi.org/10.1088/1674-1137/abce10}{\emph{Chin. Phys. C}
  {\bfseries 45} (2021) 023111}
  [\href{https://arxiv.org/abs/2003.13740}{{\ttfamily 2003.13740}}].

\bibitem{Schmidt:2018hvu}
C.~Schmidt, J.~Pumplin and C.-P.~Yuan, \emph{{Updating and optimizing error
  parton distribution function sets in the Hessian approach}},
  \href{https://doi.org/10.1103/PhysRevD.98.094005}{\emph{Phys. Rev. D}
  {\bfseries 98} (2018) 094005}
  [\href{https://arxiv.org/abs/1806.07950}{{\ttfamily 1806.07950}}].

\bibitem{Evans:2008zzb}
L.~Evans and P.~Bryant, \emph{{LHC Machine}},
  \href{https://doi.org/10.1088/1748-0221/3/08/S08001}{\emph{JINST} {\bfseries
  3} (2008) S08001}.

\bibitem{FCC:2018byv}
{\scshape FCC} collaboration, \emph{{FCC Physics Opportunities}: {Future
  Circular Collider Conceptual Design Report Volume 1}},
  \href{https://doi.org/10.1140/epjc/s10052-019-6904-3}{\emph{Eur. Phys. J. C}
  {\bfseries 79} (2019) 474}.

\bibitem{FCC:2018vvp}
{\scshape FCC} collaboration, \emph{{FCC-hh: The Hadron Collider}: {Future
  Circular Collider Conceptual Design Report Volume 3}},
  \href{https://doi.org/10.1140/epjst/e2019-900087-0}{\emph{Eur. Phys. J. ST}
  {\bfseries 228} (2019) 755}.

\bibitem{CEPCStudyGroup:2018rmc}
{\scshape CEPC Study Group} collaboration, \emph{{CEPC Conceptual Design
  Report: Volume 1 - Accelerator}},
  \href{https://arxiv.org/abs/1809.00285}{{\ttfamily 1809.00285}}.

\bibitem{Alwall:2011uj}
J.~Alwall, M.~Herquet, F.~Maltoni, O.~Mattelaer and T.~Stelzer, \emph{{MadGraph
  5 : Going Beyond}},
  \href{https://doi.org/10.1007/JHEP06(2011)128}{\emph{JHEP} {\bfseries 06}
  (2011) 128} [\href{https://arxiv.org/abs/1106.0522}{{\ttfamily 1106.0522}}].

\bibitem{Alwall:2014hca}
J.~Alwall, R.~Frederix, S.~Frixione, V.~Hirschi, F.~Maltoni, O.~Mattelaer
  et~al., \emph{{The automated computation of tree-level and next-to-leading
  order differential cross sections, and their matching to parton shower
  simulations}}, \href{https://doi.org/10.1007/JHEP07(2014)079}{\emph{JHEP}
  {\bfseries 07} (2014) 079} [\href{https://arxiv.org/abs/1405.0301}{{\ttfamily
  1405.0301}}].

\bibitem{ATLAS:2019hxz}
{ATLAS Collaboration}, \emph{{Measurements of top-quark pair differential and
  double-differential cross-sections in the $\ell$+jets channel with $pp$
  collisions at $\sqrt{s}=13$ TeV using the ATLAS detector}},
  \href{https://doi.org/10.1140/epjc/s10052-019-7525-6}{\emph{Eur. Phys. J. C}
  {\bfseries 79} (2019) 1028}
  [\href{https://arxiv.org/abs/1908.07305}{{\ttfamily 1908.07305}}].

\bibitem{Hou:2019efy}
T.-J.~Hou et~al., \emph{{New CTEQ global analysis of quantum chromodynamics
  with high-precision data from the LHC}},
  \href{https://doi.org/10.1103/PhysRevD.103.014013}{\emph{Phys. Rev. D}
  {\bfseries 103} (2021) 014013}
  [\href{https://arxiv.org/abs/1912.10053}{{\ttfamily 1912.10053}}].

\end{thebibliography}\endgroup
 
\end{document}